\documentclass[a4paper,11pt]{amsart}
\begin{document}
\hyphenation{mo-dels gra-vi-ta-tio-nal re-la-ti-vi-ty va-ria-tion
cos-mo-lo-gy}
\title[On Friedmann's universes] {{\bf  On Friedmann's universes}}
\author[Angelo Loinger]{Angelo Loinger}
\thanks{To be
published on \emph{Spacetime \& Substance.} \\email:
angelo.loinger@mi.infn.it\\ Dipartimento di Fisica, Universit\`a
di Milano, Via Celoria, 16 - 20133 Milano (Italy)}

\begin{abstract}
There is a perfect concordance between Friedmann's cosmological
models and the correspondent (purely gravitational interactions
and negligible pressure) Newtonian models. This renders quite
intuitive the fact that in general relativity no motion of bodies
generates gravitational waves.
\end{abstract}

\maketitle

\vskip1.20cm
\noindent \textbf{1}. -  As it is well known, Friedmann's
cosmological models \cite{1} can be characterized in the following
way:

\par \emph{i}) A spacetime interval $\textrm{d}s$ given by Friedmann-Robertson-Walker
 metric $(x^{1}, x^{2}, x^{3}, r$ are dimensionless coordinates):

\begin{equation} \label{eq:one}
\textrm{d}s^{2} = \left(dx^{0}\right)^{2} -
A^{2}\left(r^{2}\right)F^{2}\left(x^{0}\right)
\left[\left(\textrm{d}x^{1}\right)^{2}
    +  \left(\textrm{d}x^{2}\right)^{2} + \left(\textrm{d}x^{3}\right)^{2}\right] \quad ,
\end{equation}

where

\begin{equation} \label{eq:oneprime}
    A\left(r^{2}\right) \equiv 1 + \zeta r^{2}/4 , \tag{1'}
\end{equation}

and $\zeta=0,\pm 1$, $r^{2}= (x^{1})^{2}+(x^{2})^{2}+(x^{3})^{2}$;
$F(x^{0})$ is a function of time to be determined by Einstein
field equations. Eq.(\ref{eq:one}) yields the most general metric
expression describing a \emph{uniform} (i.e. homogeneous and
isotropic) three-dimensional space. ($\zeta=0$: Euclidean space;
$\zeta=1$: spherical -- or elliptical -- space; $\zeta=-1$:
hyperbolic space).

\par \emph{ii}) A mass tensor $T^{jk}$, $(j,k=0,1,2,3)$, of the
galaxy ``dust'', given by

\begin{equation} \label{eq:two}
    T ^{jk} = \rho \frac{\textrm{d}x^{j}}{\textrm{d}s}
    \frac{\textrm{d}x^{k}}{\textrm{d}s}, \qquad (c=1),
\end{equation}

where $\rho$ is the invariant mass density. A ``dust'' particle
represents a galaxy. The particles interact \emph{only}
gravitationally. Eq.(\ref{eq:one}) is written in a particular
\emph{Gaussian-normal} coordinate frame \cite{2} --
\emph{synchronous} frame, according to the expressive terminology
of Lan\-dau-Lifchitz \cite{2}. For frames of this kind the time
lines are geodesic lines. Remark that for \emph{any} physical
system of particles, the mass tensor of which is given by
eq.(\ref{eq:two}), the \emph{world} lines of the particles are
\emph{geodesic} lines, and therefore \emph{no} gravitational wave
can be emitted \cite{3}. Thus, our coordinate frame is synchronous
and comoving \cite{4} -- \emph{and the galaxy trajectories are}
\emph{\textbf{geodesic}} \emph{lines}. We have:

\begin{equation} \label{eq:three}
\textrm{d}x^{\alpha}/{\textrm{d}s}=0 , \quad (\alpha=1,2,3) ;
\quad \textrm{d}x^{0}/{\textrm{d}s}=1 ; \quad T^{00}=T_{00}=\rho .
\end{equation}

Substituting the preceding equations into Einstein field
equations:

\begin{equation} \label{eq:four}
    R_{jk} -\frac{1}{2} g_{jk}R = -\kappa T_{jk};  \qquad \kappa \equiv 8\pi G,
\end{equation}

where $G$ is the gravitational constant, one obtains:

\begin{equation} \label{eq:five}
    \frac{\zeta}{F} + \frac{\dot{F}^{2}}{F} + 2\ddot{F} =0,
\end{equation}

\begin{equation} \label{eq:six}
    \frac{\zeta}{F^{2}} + \frac{\dot{F}^{2}}{F^{2}} -\frac{1}{3} \kappa \rho
    =0.
\end{equation}

Clearly, the distance between two galaxies is \emph{proportional}
to the function $F$ of time. The solutions of eqs.(\ref{eq:five}),
(\ref{eq:six}) are well known: for $\zeta=1$, we have a cosmic
periodic oscillation between $F=0$ and $F=F_{max}$; for
$\zeta=0,-1$ we have a cosmic monotonic expansion. In all cases,
$\zeta/F^{2}(x^{0})$ is the curvature of the space section
$x^{0}=\textrm{constant}$. For $F=0$ the density $\rho$ is
infinite and the gravitational field is singular.

\vskip0.50cm
\noindent \textbf{2}. - In the Thirties of the past century, a
Newtonian cosmology was developed: an investigation of the kinds
of flow in a galaxy ``dust'', when homogeneity, isotropy, and
irrotational flow are assumed \cite{5}. A galaxy which is on the
surface of an expanding sphere of radius $\Re(t)$ is attracted by
the mass $M$ within the sphere according to Newton's law:

\begin{equation} \label{eq:seven}
    \ddot{\Re} = - \frac{\kappa M}{8\pi\Re^{2}}; \qquad M=
    \frac{4}{3}\pi\Re^{3}(t)\rho(t)=\textrm{constant};
\end{equation}

thus:

\begin{equation} \label{eq:sevenprime}
    \ddot{\Re} + \frac{\kappa}{6}\Re \rho=0 , \tag{7'}
\end{equation}

from which:

\begin{equation} \label{eq:eight}
    \frac{\eta}{\Re^{2}} + \frac{\dot{\Re}^{2}}{\Re^{2}} -
    \frac{1}{3}\kappa\rho=0 ,
\end{equation}

where $\eta$ is an integration constant, which characterizes the
total energy, kinetic plus potential.

\par In conclusion, we can write the two following equations:

\begin{equation} \label{eq:sevensecond}
    \frac{\eta}{\Re} + \frac{\dot{\Re}^{2}}{\Re} + 2 \ddot{\Re} =
    0 , \tag{7''}
\end{equation}

\begin{equation}  \label{eq:eightbis}
    \frac{\eta}{\Re^{2}} + \frac{\dot{\Re}^{2}}{\Re^{2}} -
   \frac{1}{3}\kappa\rho=0 . \tag{8}
\end{equation}

If we put $\eta=0,\pm1$, eqs. (7'') and (\ref{eq:eight}) become
identical to Friedmann's eqs. (\ref{eq:five}) and (\ref{eq:six}).

\par We have here three Newtonian models which, starting from a
singular point of an infinite mass density, give either a
monotonic expansion (for $\eta=0,-1$) or a periodic oscillation
(for $\eta=1$) between $\Re=0$ and $\Re=\Re_{max}$. \emph{The
concordance with Friedmann's results is complete}.

\par  The Newtonian eqs. (7'') and
(\ref{eq:eight}) give in a three-dimensional Euclidean
\emph{Bildraum} three faithful representations of the Einsteinian
eqs.(\ref{eq:five}) and (\ref{eq:six}). \emph{Vice-versa}, the
Einsteinian eqs. (\ref{eq:five}) and (\ref{eq:six}) give in three
three-dimensional \emph{Bildr\"aume} (resp. Euclidean, spherical,
hyperbolic) three faithful representations of the Newtonian eqs.
(7'') and (\ref{eq:eight}).

\vskip0.50cm
\noindent \textbf{3}. - The Newtonian images of the Einsteinian
Friedmann's models render quite intuitive the physical fact that
in general relativity \emph{no} motion of bodies generates
gravitational waves \cite{6}. (Let us observe that Birkhoff's
theorem \cite{7} regards only the \emph{outside} space of a
spherically symmetric mass distribution).

\par Another remark. The Oppenheimer-Snyder treatment of the
spherically symmetric collapse of a ``dust'' (with negligible
pressure) -- simplified image of a gravitationally collapsing star
-- utilizes Friedmann's formalism relative to the return motion in
a periodic oscillation, from $F=F_{max}$ to $F=0$ \cite{8}. These
authors describe the gravitational field in the outside space of
the star with the standard Hilbert-Droste-Weyl form of solution of
the Schwarzschild problem, improperly called ``Schwarzschild
solution''. Now, this form is actually \emph{not} suitable to the
present problem, because it is mathematically and physically valid
only for $r>2GM/c^{2}$. Neglecting erroneously this \emph{fact},
one obtains a black hole as a final product of the contraction.
But if we employ the ORIGINAL form of solution given by
Schwarzschild \cite{9}, which is regular for \emph{all} values of
the radial coordinate $r$ different from zero (and is
diffeomorphic to the exterior part, $r>2GM/c^{2}$, of the standard
HDW-form), we obtain simply a point mass, i.e. a quite usual
object \cite{10} in lieu of a fictive black hole.

\par This conclusion is corroborated by the following
consideration. If we employ the Newtonian formalism
(sect.\textbf{2}.) relative to the return motion in a periodic
oscillation, the problem of the sphero-symmetric gravitational
collapse of a star becomes quite simple -- and the final result is
quite standard: a Newtonian point mass. (Very keen papers on
gravitational collapse -- from both the standpoints, Einsteinian
and Newtonian -- have been published by Mc Vittie \cite{11}).

\par The so-called \emph{observed} black holes are only very large
-- or enormously large -- masses concentrated in relatively small
volumes. as it can be verified very easily with a careful reading
of the concerned papers.

\vskip0.50cm
\noindent \textbf{4}. - The previous formalisms are well known in
the pertinent literature, only the \emph{perspective} according to
which they are here considered is new. I have emphasized the great
significance of Friedmann's results in regard to the question of
the gravitational waves.

\par There is a perfect parallelism between the time variation of
the curvature $\zeta/F^{2}(t)$ of the spatial section
$t=\textrm{const}$ and the time variation of its Newtonian
analogue $\eta/\Re^{2}(t)$, which is proportional to Newton's
force $\kappa M/[8\pi \Re^{2}(t)]$.

\par In general -- and roughly speaking: a displacement of a given
mass causes a corresponding displacement of its Newton's force and
a corresponding displacement of the Einsteinian space curvature.
\emph{No gravitational wave is ever generated}, the analogy
between Einstein field and Maxwell-Lorentz field is a misleading
analogy.

\vskip2.0cm
\begin{center}
$^{\star---------------------------\star}$
\end{center}

\end{document}